\documentclass[conference]{IEEEtran}
\IEEEoverridecommandlockouts
\usepackage{cite}
\usepackage{amsmath,amssymb,amsfonts}
\usepackage{hyperref}
\usepackage{algorithmic}
\usepackage{tikz}
\usetikzlibrary{positioning, shapes, arrows.meta}
\usepackage{graphicx}
\usepackage{textcomp}
\usepackage{xcolor}
\usepackage{array,booktabs} 

\def\BibTeX{{\rm B\kern-.05em{\sc i\kern-.025em b}\kern-.08em
    T\kern-.1667em\lower.7ex\hbox{E}\kern-.125emX}}
\usepackage[utf8]{inputenc}
\begin{document}

\title{Investigating 1-Bit Quantization in Transformer-Based Top Tagging
\thanks{The authors acknowledge financial support from the I.I.T. Jodhpur, India, under Project No. I/RIG/JTK/20240067, and from the DST-SERB, India, under Grant No. EEQ/2023/000959.}
}

\author{\IEEEauthorblockN{Saurabh Rai}
\IEEEauthorblockA{\textit{Department of Physics} \\
\textit{Indian Institute of Technology Kanpur}\\
Kanpur, India \\
saurabhrai25@iitk.ac.in}
\and
\IEEEauthorblockN{Prisha}
\IEEEauthorblockA{\textit{Department of Physics} \\
\textit{Indian Institute of Technology Jodhpur}\\
Jodhpur, India \\
p23ph0008@iitj.ac.in}
\and
\IEEEauthorblockN{Jitendra Kumar}
\IEEEauthorblockA{\textit{Department of Physics} \\
\textit{Indian Institute of Technology Jodhpur}\\
Jodhpur, India \\
jkumar@iitj.ac.in}
}

\maketitle

\begin{abstract}
The increasing scale of deep learning models in high-energy physics (HEP) has posed challenges to their deployment on low-power, latency-sensitive platforms, such as FPGAs and ASICs used in trigger systems, as well as in offline data reconstruction and processing pipelines. In this work, we introduce BitParT, a 1-bit Transformer-based architecture designed specifically for the top-quark tagging method. Building upon recent advances in ultra-low-bit large language models (LLMs), we extended these ideas to the HEP domain by developing a binary-weight variant (BitParT) of the Particle Transformer (ParT) model. Our findings indicate a potential for substantial reduction in model size and computational complexity, while maintaining high tagging performance. We benchmark BitParT on the public Top Quark Tagging Reference Dataset and show that it achieves competitive performance relative to its full-precision counterpart. This work demonstrates the design of extreme quantized models for physics applications, paving the way for real-time inference in collider experiments with minimal and optimized resource usage.
\end{abstract}

\begin{IEEEkeywords}
Top-quark tagging, Large Hadron Collider, 1-bit LLMs, Transformers, Efficient Deep Learning
\end{IEEEkeywords}

\section{Introduction}
Particle physics research aims to explore the universe at the most fundamental level. The Standard Model (SM) of particle physics is a remarkably successful framework for describing the fundamental building blocks of nature and the forces that govern their interactions. Yet, many open questions remain, from the matter-antimatter asymmetry to the mystery of dark matter. 

High energy accelerators are used as powerful tools to mimic the conditions of the early universe by colliding particles at ultra-relativistic velocities. For example, large-scale experiments like CMS and ATLAS have been built at the Large Hadron Collider (LHC), where protons are accelerated nearly the speed of light and collide at unprecedented energies in the multi-TeV range. These collisions occur at an extremely high rate, and the outcome of each collision recorded by the detectors is defined as an ``event", capturing key information such as particle trajectories, momentum, and energies at the detector level. A collection of recorded events is referred to as a dataset and is used for physics measurements.

In the SM, quarks and leptons are considered as fundamental constituents of matter. The visible matter in the universe is mostly made up of the lightest quarks, up and down, along with electrons, all bound together in atoms. The strange quark, while also considered light, usually appears only in unstable hadrons. The heavier quarks, charm and bottom, decay rapidly and are found in unstable hadrons as well. The \textit{top quark} mass is 172.76 $\pm$ 0.3 GeV/c$^{2}$, making it the heaviest. It decays almost instantly due to its very short mean lifetime of $5 \times 10^{-25}$ seconds, and is unique as it does not form hadrons at all.

Because the top quark decays before hadronization, it allows probing the bare quark properties and performing precision Quantum Chromodynamics (QCD) studies. It also couples strongly to the Higgs field, influencing electroweak symmetry breaking and providing a sensitive probe for physics beyond the Standard Model (BSM). The decay products of the top quark include quarks and gluons that rapidly hadronize, forming collimated streams of particles known as jets. This is where ``\textit{top tagging}" becomes important, as it enables the identification of jets originating from top quark decays, distinguishing them from those arising from other processes. It leverages jet reconstruction algorithms to analyze the substructure of the jets and infer the presence of a top quark based on characteristic decay patterns. 

The conventional top tagging algorithm used at the Tevatron relied on b‑tagging, where top quarks were produced near threshold, so their decay products, especially the bottom quark jets, were distinctly separated and could be distinguished using secondary vertex or soft-lepton b-tagging, combined with conventional jet reconstruction and kinematic selection \cite{CDF1995, D0a1995, D0b1995}. At the LHC, however, top quarks are often boosted, so advanced methods such as $k_T$ and anti‑$k_T$ algorithms with large‑R jets, subjet b‑tagging, substructure variables, and grooming techniques are used \cite{thaler2008, bKaplan2008, abdesselam2011, plehn2012,  altheimer2012, altheimer2014, bAlmeida2009a,bAlmeida2009b,Larkoski2017jix}. These methods exploit the internal structure and kinematic patterns of the top quark decay products. Although these methods perform well for top tagging, they face performance challenges under extreme pile-up and suffer from large systematic uncertainties. 

Machine learning (ML) techniques bring several major advances to the field of top tagging. Instead of relying only on jet substructure variables, ML models can work directly with low-level jet information, exploiting complex and nonlinear correlations between jet observables\cite{Guest:2018yhq, bInterplay2024, Bhattacharya:2020aid, Larkoski2017jix}. Consequently, a range of approaches have been designed for ML-based top tagging, starting from boosted decision trees (BDTs) and convolutional neural networks (CNNs) \cite{bCMSBTV162} and fully connected dense neural networks\cite{deOliveira:2015xxd, ATLAS:2017bqn, atlas2019topwtagging}, to more advanced graph neural networks (GNNs) \cite{Dreyer:2020brq, Gong:2022lye, Ma:2022bvt}, point cloud networks \cite{Qu:2019gqs, Mikuni:2021pou}, and recently transformer models \cite{Mikuni:2021pou, Qu:2022mxj, He:2023cfc, Wu:2024thh, Brehmer:2024yqw, Wang:2024rup, Bardhan:2025icr}. These ML methods not only served as an efficient choice but also pushed tagging performance forward. Despite all these advancements, the enormous size and complexity of the LHC data still impose enormous computing demands, both in offline data reconstruction and at the trigger level. The upcoming High-Luminosity LHC (HL-LHC) will add additional challenges due to higher collision rates, larger data volumes, and processing requirements.

Deep learning architectures for LHC data processing often require large and complex models to meet the demands of high data volume and complexity. They also rely on extensive training datasets and require fast, high-throughput inference. All of this makes them challenging to scale and deploy efficiently on available computing  \cite{Guest:2018yhq, Iiyama:2020wap}. Similar requirements are also common in computer science and the machine learning domain when deploying DL models on mobile, embedded, or resource-limited platforms. To address this, the broader community has developed a range of model compression techniques, including quantization \cite{Hubara:2018hmd, Han:2015, bal2024spiking}, pruning \cite{Denton:2014}, and low-rank approximations \cite{Jaderberg:2014}, which not only reduce memory but also compute costs while preserving task performance. A notable recent advancement in this direction is the BitNet family of architectures \cite{BitNet:2023, 1.58bit:2024, ma2024bitnet, nielsen2024bitnet}, replacing full-precision floating-point multiplications with low-bit integer operations that can be efficiently implemented using bitwise computation. The BitNet approach to extreme low-bit quantization aims to maximize computational and energy efficiency while maintaining competitive performance, offering a promising direction for scalable, low-power deep learning model integration into HEP experiment pipelines.

Motivated by these recent developments, we explore a 1-bit quantized variant of the Particle Transformer (\textbf{\texttt{ParT}}) architecture, which we refer to as the \textbf{\texttt{BitParT}} model, as a case study for top quark tagging. Our work, based on the BitNet approach for extreme quantization, addresses a similar direction as a recent study in the HEP community \cite{Krause:2025qnl}, which demonstrated the viability of low-bit quantization across various tasks. However, the two methods differ in focus and design of \texttt{BitLinear} layers. Our results show that BitParT delivers similarly high performance to its full-precision counterpart for top-tagging, offering not only a viable deployment option for resource-constrained experimental contexts but also an alternative machine learning approach.

\section{Dataset and Selections}
We used the top quark tagging reference dataset established by Butter et al. in \cite{Butter2019}. This dataset contains 2 $\times$ 10$^6$ proton-proton collision events generated with the Pythia8 generator at a center-of-mass energy of 14 TeV. The data comes with an equal proportion of signal (top quark jets) events and underlying background events (mixed quark-gluon jets). These events are produced without involving multiple interactions or pile-up, and the detector-level simulation is performed using the Delphes ATLAS detector card. 

The reconstruction of the jets is performed using the anti-$k_T$ algorithm with radius parameter $R=0.8$~\cite{FastJet2012}. The other selection criteria are specified in \cite{Butter2019}. For example, for each event, only the leading jet is considered for which the transverse momentum is restricted to the range 550-650 GeV and pseudorapidity $|\eta| < 2.0$, corresponding to the highly boosted regime where top decay products become collimated within a single large-radius jet. The signal jet is matched to parton-level top quarks within $\Delta R=0.8$ and additionally requires all top decay products ($b$ quark and $W$ boson decay) to lie within the jet cone. For particle-level input representation, each jet is represented as an unordered set of the leading 200 constituent particles, and a jet containing fewer particles is zero-padded to maintain consistent input dimensions. 

The jet constituent particles correspond to particle flow objects from the Delphes reconstruction. Particles are then ordered by transverse momentum ($p_{T}$) with the highest-$p_{T}$ constituent ranked first. Each particle is described by features drawn from three categories: kinematic properties, particle identification variables, and trajectory displacement parameters, as detailed in \cite{Qu:2022mxj}. Since we are using a simulated dataset, the truth information for the top quark four-momentum is also available and used to classify the signal events and assign binary labels. Finally, for our ML study, we divided the input signal and background dataset into three categories: training, validation, and testing, with sample proportions of 60\%, 20\%, and 20\%, respectively.

\begin{table}[h]
\centering
\small
\caption{Particle Level Kinematic variables~\cite{Qu:2022mxj}.}
\setlength{\tabcolsep}{1pt} 
\renewcommand{\arraystretch}{1.1} 
\begin{tabular}{c>{\centering\arraybackslash}m{2.8cm}m{4.9cm}}
\toprule
\textbf{S.No.} & \textbf{Variable} & \textbf{Description} \\
\midrule
1 & $\Delta \eta$, $\Delta \phi$ & Pseudorapidity and azimuthal angle differences between the particle and the jet axis \\
2 & $\log p_{T}$, $\log E$ & Logarithm of the particle’s transverse momentum $p_{T}$ and energy $E$\\
3 & $\log \frac{p_T}{p_T^{\mathrm{jet}}},\ \log \frac{E}{E^{\mathrm{jet}}}$ & Logarithms of the particle’s $p_T$ and $E$ relative to the jet $p_T$ and $E$ \\
4 & $\Delta R$ & Angular distance in $(\eta, \phi)$ space between the particle and the jet axis; ($\sqrt{(\Delta \eta)^{2} + (\Delta \phi)^{2}}$) \\
\bottomrule
\end{tabular}

\label{tab:kinematics}
\end{table}

\section{Model Architectures}
\subsection{\textbf{Particle Transformer (ParT) Architecture}}
The Particle Transformer (ParT) is a jet tagging model based on the Transformer architecture, specifically adapted to handle variable-length sets of jet constituents \cite{Qu:2022mxj}. In the model, each jet is treated as an unordered set of particles, with each particle described by a rich set of features. To better capture the jet’s internal structure, interactions between particles are also explicitly encoded as paired particle features. The full model architecture is described in \cite{Qu:2022mxj}; here, we highlight only the key components and our specific optimizations for BitPart.

\noindent\textbf{Input Representation:}  
For each jet, the inputs are composed of two components: a matrix \( X \in \mathbb{R}^{N \times d} \), which contains per-particle $C$ features for each of the \( N \) particles in a jet, and a tensor \( U \in \mathbb{R}^{N \times N \times d'} \), which encodes pairwise $C^\prime$ features from relationships between particles. Here, \( d \) and \( d' \) denote the dimensions of the particle and pairwise feature embeddings, respectively.

\noindent\textbf{Particle and Pairwise Interaction Encoding:}
The particle-level features include the kinematic variables listed in Table~\ref{tab:kinematics}. These are projected via a multi-layer perceptron (MLP) into a $d$-dimensional embedding space. To account for correlations between particle pairs, a symmetric interaction tensor is constructed using observables derived from the four-momenta of the particles, and in principle, it can incorporate any relevant interaction features. For any pair of particles $i$ and $j$, following are defined:

\[
\begin{aligned}
\Delta R_{ij} &= \sqrt{(y_i - y_j)^2 + (\phi_i - \phi_j)^2}, \\
k_{T,ij} &= \min(p_{T,i}, p_{T,j}) \cdot \Delta R_{ij}, \\
z_{ij} &= \frac{\min(p_{T,i}, p_{T,j})}{p_{T,i} + p_{T,j}}, \\
m_{ij}^2 &= (E_i + E_j)^2 - \|\vec{p}_i + \vec{p}_j\|^2
\end{aligned}
\]

Here, $y$ denotes the rapidity, $\phi$ represents the azimuthal angle and the logarithms of these quantities, i.e., $\log \Delta R$, $\log k_T$, $\log z$, and $\log m^2$, are used as the interaction features and subsequently processed through a shared 1D convolutional encoder to yield the final attention bias tensor $U$. The choice of the logarithmic quantities is because the distributions show high-side tails, and is adopted with the approach used in the previous work ~\cite{Dreyer:2020brq,  Qu:2022mxj}.

\noindent\textbf{Modified Self-Attention:}  
The core building block of ParT is a modified self-attention mechanism where pairwise interaction biases are injected into the attention computation. Specifically, for query-key-value triplets $(Q, K, V)$ derived from the particle embeddings, attention is computed as:
\[
\text{Attention}(Q, K, V) = \text{Softmax}\left( \frac{QK^\top}{\sqrt{d_k}} + U \right) V,
\]
where $U$ is broadcast appropriately to match the number of attention heads. This formulation allows the model to directly modulate attention weights based on physics-informed features of particle pairs.

\noindent\textbf{Model Structure:}  
The architecture is organized into two stages. The first stage consists of a stack of $L$ particle attention blocks, where each block includes the modified multi-head attention followed by a position-wise feedforward MLP, both wrapped with residual connections and normalization layers. Notably, no positional encodings are used, preserving permutation invariance. In the second stage, a special learnable class token is introduced and allowed to attend to the set of particle embeddings via two class attention blocks. These blocks use standard attention (without pairwise biases) to aggregate information into the class token, which is then passed through a final classification head.

\begin{figure}[htbp]
    \centering
    \includegraphics[scale=0.85]{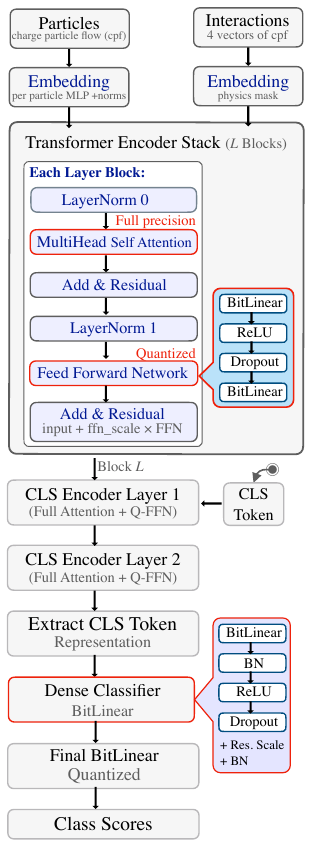}
    \caption{The Architecture of BitPart Model}
    \label{bitpartmodel}
\end{figure}

\subsection{\textbf{1-bit Particle Transformer (BitParT) Architecture}}

The \textit{BitParT} is a quantized variant of the ParT, designed to reduce model size and computational cost through selective 1-bit quantization, while preserving components essential for modeling physics-specific structure. It integrates a custom binary linear layer called \texttt{BitLinear}, which replaces standard linear projections in the feed-forward networks and classifier, but retains full-precision computation in attention mechanisms and physics-informed embedding modules.

\noindent\textbf{BitLinear Quantization:}
Given an input matrix $x \in \mathbb{R}^{B \times D}$ and a weight matrix $W \in \mathbb{R}^{D_{\text{out}} \times D}$, the \texttt{BitLinear} layer applies a two-step quantization:
\begin{enumerate}
    \item \textit{Input Quantization}: Each input vector is mean-centered per row, $\tilde{x} = x - \mu_x$ with $\mu_x = \mathrm{mean}(x, \mathrm{dim}=-1)$, binarized as $\mathrm{sign}(\tilde{x}) \in \{-1, +1\}^D$, and rescaled by $\alpha_x = \mathrm{mean}(|\tilde{x}|)$ to ensures magnitude preservation.   
    \item \textit{Weight Quantization}: The weights are quantized as $\mathrm{sign}(W) \in \{-1, +1\}$ after mean-centering, with a per-batch scaling factor $\alpha_w = \mathrm{mean}(|a|)$ to preserve dynamic range. An optional learnable scalar $\alpha_{\text{learnable}}$ is applied to adapt the scaling during training.
\end{enumerate}
The final output is computed as:
\[
y = \alpha_x \cdot \alpha_w \cdot \alpha_{\text{learnable}} \cdot (\text{sign}(\tilde{x}) W_{\text{sign}}^T) + b,
\]
where $b$ is the optional bias vector and the matrix product is performed using bitwise operations. Gradients are backpropagated using the straight-through estimator (STE), i.e., $\nabla \text{sign}(x) \approx \nabla x$.

\noindent\textbf{BitLinLayer Block:} To ensure stable training under quantization, each quantized projection is wrapped in a \texttt{BitLinLayer} block composed of:  
\begin{itemize}
    \item \texttt{BitLinear} projection,  
    \item First Batch normalization (BN1),  
    \item ReLU activation,  
    \item Dropout layer,  
    \item Residual connection with learnable scaling,\\ $x+\beta\cdot\mathrm{Block}(x)$,  
    \item Second batch normalization (BN2) applied after the residual merge.  
\end{itemize}

\noindent\textbf{Transformer Encoder with Quantized Feed-forward Network (FFN):} The transformer layers retain the original structure but apply quantization only to the feed-forward network:  
\begin{itemize}
    \item \textit{Attention:} Each encoder layer contains a standard \texttt{MultiheadAttention} module operating in full precision. A residual connection followed by layer normalization is applied.
    \item \textit{Feed-Forward Network (FFN):} Implemented as 
    \(
    x \rightarrow \mathrm{BitLinear}(d \!\rightarrow\! 4d) \rightarrow \mathrm{ReLU} \rightarrow \mathrm{Dropout} \rightarrow \mathrm{BitLinear}(4d \!\rightarrow\! d)
    \),  
    with output scaled by a learnable factor $\gamma_{\mathrm{ffn}}$ before being added back via a residual connection and normalized.
\end{itemize}

\noindent\textbf{CLS Attention Mechanism:}
A pair of specialized transformer blocks (\texttt{BitCLS\_TransformerEncoderLayer}) performs classification token pooling using full-precision self-attention over the CLS token concatenated with encoded particles. The feed-forward layers are quantized similarly to the standard encoder, with layer-specific scaling and normalization.

\noindent\textbf{Classifier:}
The CLS token output is passed through a small quantized dense classifier composed of a single \texttt{BitLinLayer}, followed by a final \texttt{BitLinear} projection to the number of output classes.

\noindent\textbf{Unquantized Components:}
To maintain sensitivity to physically meaningful correlations, the following components remain in full precision:
\begin{itemize}
    \item Attention blocks: All \texttt{MultiheadAttention} operations,
    \item Layer Normalization,
    \item Pairwise Mask Embedding (\texttt{PairEmbed}) module,
    \item Input Processing (e.g., per-particle MLPs),
    \item CLS Token: Learnable vector with standard initialization and precision.
\end{itemize}
The BitParT architecture reduces precision in a controlled and targeted manner, focusing quantization on high-parameter-cost modules (FFNs and classifier) while retaining the inductive priors encoded in attention and physics-aware modules. This enables substantial model compression and potential for efficient deployment, while preserving physics fidelity and performance on jet tagging tasks.
\section{Performance Analysis}

To evaluate the effectiveness of binary quantization under realistic physics conditions, we compare a custom full-precision \texttt{ParT} architecture with its quantized variant, \texttt{BitParT}, on a top-quark tagging task. Both models used in our study are architecturally identical, containing approximately 1.23M trainable parameters, thereby ensuring that performance differences are attributable solely to quantization effects rather than architectural discrepancies. For comparison, we use three different metrics, namely, classification accuracy, area under the receiver operator characteristic curve (AUC) and the background rejection at a fixed signal efficiency.

The models employ 3 Transformer blocks with 8 attention heads per block, an embedding dimensionality of 128, and a feed-forward hidden dimension of 512. These blocks are followed by two additional Transformer layers dedicated to refining a class token (\texttt{CLS}), which interacts with the sequence output via multi-head self-attention. This CLS token serves as the basis for final classification. Training is conducted using the AdamW optimizer with an initial learning rate of $10^{-3}$, batch size of 512, and weight decay of 0.01. A cosine annealing learning rate schedule was used to progressively reduce the learning rate, reaching 1\% of its initial value over the final 30\% of the training epochs. Both the full-precision and quantized models were trained for 20 epochs.

\noindent\textbf{Quantization Scope:}
The \texttt{BitParT} model applies strict 1-bit quantization to all weights within the feed-forward networks and the dense classifier layers using the custom \texttt{BitLinear} module. Attention mechanisms, input embedding modules, normalization layers, and all bias parameters remain in full precision. Analysis of parameter allocation reveals that over 67\% of the model’s weights are subject to 1-bit quantization, ensuring substantial compression while preserving critical inductive biases in full-precision components. Gradient flow through quantized paths is enabled via quantization-aware training with straight-through estimators.

\begin{figure}[htbp]
\centerline{\includegraphics[width=0.5\textwidth]{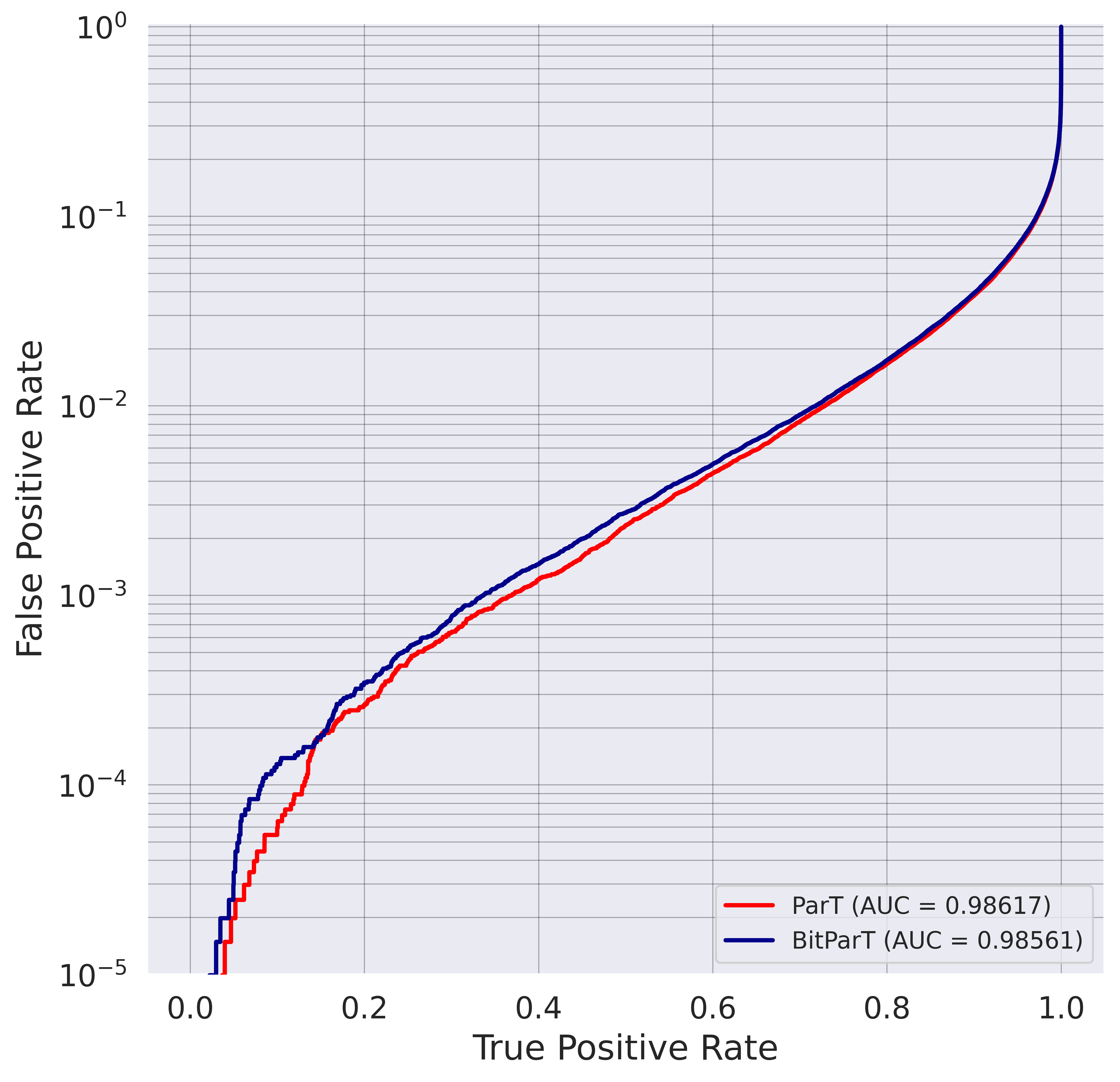}}
\caption{Comparison of AUC scores for ParT and BitParT models. }
\label{fig2}
\end{figure}

\noindent\textbf{Classification Accuracy and AUC:}
Despite aggressive quantization, \texttt{BitParT} achieves an accuracy of 93.99\% against the benchmark accuracy of 94\% of \texttt{ParT} on the test set. Receiver Operating Characteristic (ROC) analysis shows that the full-precision model attains an AUC of 0.9862, while the quantized variant reaches 0.9856 (Fig. \ref{fig2}). The relative difference of less than 0.15\% is well within the statistical uncertainties and negligible in the context of experimental systematic uncertainties. These results validate the robustness of binary quantization for preserving the overall discriminative power in top tagging applications.

\noindent\textbf{Background Rejection Performance.}
The performance metric most relevant for collider physics is background rejection at fixed signal efficiency, as it directly correlates with experimental sensitivity in searches and measurements. The metric is defined as:
\begin{equation}
    Rejection_{X\%} = 1/FPR \hspace{0.2cm} at \hspace{0.2cm} TPR=X\%,
\end{equation}
where FPR and TPR are the false positive and true positive rates, respectively.

At 50\% signal efficiency:
\begin{itemize}
    \item \texttt{ParT} achieves a background rejection factor of $429 \pm 20$.
    \item \texttt{BitParT} yields a background rejection of $366 \pm 17$.
\end{itemize}
The observed $15\%$ reduction in rejection, while measurable, is likely to remain within acceptable limits for many practical cases, particularly in real-time inference contexts such as Level-1 trigger systems, where latency and power constraints are critical. The overlapping uncertainty intervals further suggest that the difference, though systematic, is moderate relative to typical detector-level systematic effects for a top-tagging-like task. The overall performance details of BitParT and other models (\cite{Qu:2019gqs, Mikuni:2021pou, Qu:2022mxj, Gong:2022lye}) are presented in Table \ref{tab:model_comparison}.

In summary, the \texttt{BitParT} model demonstrates that high-fidelity classification can be retained under extreme quantization, with minimal impact on AUC and manageable degradation in background rejection. Given that over 67\% of the model parameters are binarized, this approach offers a significant reduction in memory and compute requirements. Future hardware-level deployment may employ such compression strategies for efficient performance of state-of-the-art machine learning algorithms in high-energy physics.

\begin{table}[htbp]
\caption{Top tagging performance comparison for various models with reference dataset}
\begin{center}
\renewcommand{\arraystretch}{1.2}
\begin{tabular}{lccll}
\hline
Model & Accuracy & AUC & Rej$_{50\%}$ & Rej$_{30\%}$ \\
\hline
P-CNN            & 0.930 & 0.9803 & $201 \pm 4$   & $759 \pm 24$ \\
ParticleNet      & 0.940 & 0.9858 & $397 \pm 7$   & $1615 \pm 93$ \\
PCT              & 0.940 & 0.9855 & $392 \pm 7$   & $1533 \pm 101$ \\
LorentzNet       & 0.942 & 0.9868 & $498 \pm 18$  & $2195 \pm 173$ \\
ParticleNet-f.t. & 0.942 & 0.9866 & $487 \pm 9$   & $1771 \pm 80$ \\
ParT-f.t.        & 0.944 & 0.9877 & $691 \pm 15$  & $2766 \pm 130$ \\
ParT             & 0.940 & 0.9858 & $413 \pm 16$  & $1602 \pm 81$ \\\hline
ParT (Custom)             & 0.940 & 0.9862 & $429 \pm 20$  & $1570 \pm 130$ \\
\textbf{BitParT}        & \textbf{0.9399} & \textbf{0.9856} & \textbf{366 $\pm$ 17}  & \textbf{1301 $\pm$ 106} \\
\hline
\end{tabular}
\label{tab:model_comparison}
\end{center}
\end{table}

\section{Conclusion}
We have introduced the BitParT architecture, a 1-bit quantized version of the ParT model, designed for top tagging.

Despite the extreme quantization, BitParT preserves classification performance, achieving near-identical accuracy and AUC compared to the full-precision ParT with only modest reductions in background rejection at different fixed signal efficiencies. These results demonstrate that a highly compressed binary transformer model can be trained to effectively replicate the performance of its full-precision counterpart.

Although this work has not yet quantified inference latency or energy consumption improvements, the architectural design of BitParT lays the foundation for future hardware-aware optimizations. In particular, targeted FPGA or low-power inference benchmarking would provide further insight into the practical benefits of quantization in collider trigger pipelines. Additionally, extending BitParT to more diverse classification tasks, such as those present in the JetClass dataset, will allow evaluation of the model's generalization capability beyond top tagging. Future studies may also explore hybrid precision strategies and structured pruning to further improve efficiency-accuracy trade-offs in real-time jet classification settings.


\begin{thebibliography}{00}
\bibitem{CDF1995}
CDF Collaboration, ``Observation of top quark production in $p\bar{p}$ collisions'', Phys. Rev. Lett. \textbf{74}, 2626 (1995).

\bibitem{D0a1995}
D\O{} Collaboration, ``Search for high mass top quark production in $p\bar{p}$ collisions at $\sqrt{s} = 1.8$ TeV'', Phys. Rev. Lett. \textbf{74}, 2422 (1995).

\bibitem{D0b1995}
D\O{} Collaboration, ``Observation of the Top Quark", Phys. Rev. Lett. \textbf{74}, 2632 (1995).

\bibitem{thaler2008} J. Thaler, L.-T. Wang, ``Strategies to identify boosted tops". JHEP \textbf{07}, 092 (2008).


\bibitem{bKaplan2008}
D.~E.~Kaplan, K.~Rehermann, M.~D.~Schwartz and B.~Tweedie,
``Top Tagging: A Method for Identifying Boosted Hadronically Decaying Top Quarks,''
Phys. Rev. Lett. \textbf{101} (2008), 142001.

\bibitem{abdesselam2011}
A.~Abdesselam, E.~B.~Kuutmann, U.~Bitenc, G.~Brooijmans, J.~Butterworth, P.~Bruckman de Renstrom, D.~Buarque Franzosi, R.~Buckingham, B.~Chapleau and M.~Dasgupta, \textit{et al.}
``Boosted Objects: A Probe of Beyond the Standard Model Physics,''
Eur. Phys. J. C \textbf{71} (2011), 1661.

\bibitem{plehn2012}
T.~Plehn and M.~Spannowsky,
``Top Tagging,''
J. Phys. G \textbf{39} (2012), 083001.

\bibitem{altheimer2012}
A. Altheimer \textit{et al.}, ``Jet Substructure at the Tevatron and LHC: New results, new tools, new benchmarks'', J. Phys. G \textbf{39} (2012) 063001.

\bibitem{altheimer2014}
A. Altheimer \textit{et al.}, ``Boosted objects and jet substructure at the LHC: Report of BOOST2012", Eur. Phys. J. C \textbf{74} (2014) 2792.

\bibitem{bAlmeida2009a}
L.~G.~Almeida, S.~J.~Lee, G.~Perez, G.~F.~Sterman, I.~Sung and J.~Virzi,
``Substructure of high-$p_T$ Jets at the LHC,''
Phys. Rev. D \textbf{79} (2009), 074017.

\bibitem{bAlmeida2009b}
L.~G.~Almeida, S.~J.~Lee, G.~Perez, I.~Sung and J.~Virzi,
``Top Jets at the LHC,''
Phys. Rev. D \textbf{79} (2009), 074012.

\bibitem{Larkoski2017jix}
A.~J.~Larkoski, I.~Moult and B.~Nachman,
``Jet Substructure at the Large Hadron Collider: A Review of Recent Advances in Theory and Machine Learning,''
Phys. Rept. \textbf{841} (2020), 1-63.

\bibitem{Guest:2018yhq}
D.~Guest, K.~Cranmer and D.~Whiteson,
``Deep Learning and Its Application to LHC Physics,''
Ann. Rev. Nucl. Part. Sci. \textbf{68} (2018), 161-181

\bibitem{bInterplay2024}
C.~Bose, A.~Chakraborty, S.~Chowdhury and S.~Dutta,
``Interplay of traditional methods and machine learning algorithms for tagging boosted objects,''
Eur. Phys. J. ST \textbf{233} (2024) no.15-16, 2531-2558.

\bibitem{Bhattacharya:2020aid}
S.~Bhattacharya, M.~Guchait and A.~H.~Vijay, ``Boosted top quark tagging and polarization measurement using machine learning,'' Phys. Rev. D \textbf{105} (2022) no.4, 042005

\bibitem{bCMSBTV162}
A.~M.~Sirunyan \textit{et al.} [CMS],
``Identification of heavy-flavour jets with the CMS detector in pp collisions at 13 TeV,''
JINST \textbf{13} (2018) no.05, P05011.

\bibitem{deOliveira:2015xxd}
L.~de Oliveira, M.~Kagan, L.~Mackey, B.~Nachman and A.~Schwartzman,
``Jet-images {\textemdash} deep learning edition,''
JHEP \textbf{07} (2016), 069.

\bibitem{ATLAS:2017bqn}
 ATLAS Collaboration,
``Identification of Hadronically-Decaying W Bosons and Top Quarks Using High-Level Features as Input to Boosted Decision Trees and Deep Neural Networks in ATLAS at $\sqrt{s}$ = 13 TeV,''
ATL-PHYS-PUB-2017-004.

\bibitem{atlas2019topwtagging}
M.~Aaboud \textit{et al.} [ATLAS],
``Performance of top-quark and $W$-boson tagging with ATLAS in Run 2 of the LHC,''
Eur. Phys. J. C \textbf{79} (2019) no.5, 375.

\bibitem{Dreyer:2020brq}
F.~A.~Dreyer and H.~Qu,
``Jet tagging in the Lund plane with graph networks,''
JHEP \textbf{03} (2021), 052.

\bibitem{Gong:2022lye}
S.~Gong, Q.~Meng, J.~Zhang, H.~Qu, C.~Li, S.~Qian, W.~Du, Z.~M.~Ma and T.~Y.~Liu,
``An efficient Lorentz equivariant graph neural network for jet tagging,''
JHEP \textbf{07} (2022), 030.

\bibitem{Ma:2022bvt}
F.~Ma, F.~Liu and W.~Li,
``Jet tagging algorithm of graph network with Haar pooling message passing,''
Phys. Rev. D \textbf{108} (2023) no.7, 072007.

\bibitem{Qu:2019gqs}
H.~Qu and L.~Gouskos,
``ParticleNet: Jet Tagging via Particle Clouds,'' Phys. Rev. D \textbf{101} (2020) no.5, 056019.

\bibitem{Mikuni:2021pou}
V.~Mikuni and F.~Canelli,
``Point cloud transformers applied to collider physics,''
Mach. Learn. Sci. Tech. \textbf{2} (2021) no.3, 035027.

\bibitem{Qu:2022mxj}
H.~Qu, C.~Li and S.~Qian,
``Particle Transformer for Jet Tagging,''
[arXiv:2202.03772 [hep-ph]].

\bibitem{He:2023cfc}
M.~He and D.~Wang,
``Quark/gluon discrimination and top tagging with dual attention transformer,''
Eur. Phys. J. C \textbf{83} (2023) no.12, 1116.

\bibitem{Wu:2024thh}
Y.~Wu, K.~Wang, C.~Li, H.~Qu and J.~Zhu,
``Jet tagging with more-interaction particle transformer,''
Chin. Phys. C \textbf{49} (2025) no.1, 013110.

\bibitem{Brehmer:2024yqw}
J.~Brehmer, V.~Bres{\'o}, P.~de Haan, T.~Plehn, H.~Qu, J.~Spinner and J.~Thaler,
``A Lorentz-Equivariant Transformer for All of the LHC,'' [arXiv:2411.00446 [hep-ph]].

\bibitem{Wang:2024rup}
A.~Wang, A.~Gandrakota, J.~Ngadiuba, V.~Sahu, P.~Bhatnagar, E.~E.~Khoda and J.~Duarte,
``Interpreting Transformers for Jet Tagging,'' [arXiv:2412.03673 [hep-ph]].

\bibitem{Bardhan:2025icr}
J.~Bardhan, R.~Agrawal, A.~Tilak, C.~Neeraj and S.~Mitra,
``HEP-JEPA: A foundation model for collider physics using joint embedding predictive architecture,'' [arXiv:2502.03933 [cs.LG]].

\bibitem{Iiyama:2020wap}
Y.~Iiyama, G.~Cerminara, A.~Gupta, J.~Kieseler, V.~Loncar, M.~Pierini, S.~R.~Qasim, M.~Rieger, S.~Summers and G.~Van Onsem, \textit{et al.}
``Distance-Weighted Graph Neural Networks on FPGAs for Real-Time Particle Reconstruction in High Energy Physics,''
Front. Big Data \textbf{3} (2020), 598927.

\bibitem{Hubara:2018hmd}
I.~Hubara, M.~Courbariaux, D.~Soudry, R.~El-Yaniv, Y.~Bengio
``Quantized neural networks: Training neural networks with low precision weights and activations,''
Journal of Machine Learning Research \textbf{18}, no. 187 (2018): 1-30.

\bibitem{Han:2015}
S.~Han, J.~Pool, J.~Tran, W.~Dally,  
``Learning both weights and connections for efficient neural networks,''  
Advances in Neural Information Processing Systems, \textbf{28}, (2015).

\bibitem{bal2024spiking}
Malyaban Bal, Yi Jiang, Abhronil Sengupta, 
``Exploring Extreme Quantization in Spiking Language Models'', arXiv:2405.02543v2, 2024.

\bibitem{Denton:2014}
E.~L.~Denton, W.~Zaremba, J.~Bruna, Y.~LeCun, R.~Fergus,  
``Exploiting linear structure within convolutional networks for efficient evaluation,'' Advances in Neural Information Processing Systems \textbf{27} (2014).

\bibitem{Jaderberg:2014}
M.~Jaderberg, A.~Vedaldi, A.~Zisserman,  
``Speeding up convolutional neural networks with low rank expansions,''  
Proceedings of the British Machine Vision Conference (2014).

\bibitem{BitNet:2023}
H.~Wang, S.~Ma, L.~Dong, S.~Huang, H.~Wang, L.~Ma, F.~Yang, R.~Wang, Y.~Wu, F.~Wei, ``BitNet: Scaling 1-Bit Transformers for Large Language Models,''  
arXiv:2305.18245 [cs.LG].

\bibitem{1.58bit:2024}
J.~Zhou, H.~Li, Y.~Liu, Y.~Li, Y.~Jiang, C.~Zhang, H.~Zhang, W.~Lin,  
``The Era of 1-Bit LLMs: All Large Language Models are in 1.58 Bits,''  
arXiv:2403.18186 [cs.LG].

\bibitem{ma2024bitnet}
Shuming Ma, Hongyu Wang, Lingxiao Ma, Lei Wang, Wenhui Wang, Shaohan Huang, Li Dong, Ruiping Wang, Jilong Xue, Furu Wei,  
``The Era of 1-bit LLMs: All Large Language Models are in 1.58 Bits'', arXiv:2402.17764, 2024.

\bibitem{nielsen2024bitnet}
Jacob Nielsen, Peter Schneider-Kamp,  
``BitNet b1.58 Reloaded: State-of-the-art Performance Also on Smaller Networks'', arXiv:2407.09527, 2024.

\bibitem{Krause:2025qnl}
C.~Krause, D.~Wang and R.~Winterhalder,
``BitHEP -- The Limits of Low-Precision ML in HEP,'' [arXiv:2504.03387 [hep-ph]].

\bibitem{Butter2019}
G.~Kasieczka, T.~Plehn, A.~Butter, K.~Cranmer, D.~Debnath, B.~M.~Dillon, M.~Fairbairn, D.~A.~Faroughy, W.~Fedorko and C.~Gay, \textit{et al.}
``The Machine Learning landscape of top taggers,''
SciPost Phys. \textbf{7} (2019), 014.

\bibitem{FastJet2012}
M.~Cacciari, G.~P.~Salam and G.~Soyez,
``FastJet User Manual,''
Eur. Phys. J. C \textbf{72} (2012), 1896.

\end{thebibliography}
\end{document}